\documentclass[reprint,showpacs,amsmath,amssymb,floatfix,apl,superscriptaddress,showkeys,aip]{revtex4-1}

\usepackage{graphicx}
\usepackage{epsfig}
\usepackage{dcolumn}
\usepackage{bm}
\usepackage{times}

\begin{document}


\title{Quantum theory of space charge limited current in solids}

\author{Gabriel Gonz\'alez}
\email{gabriel.gonzalez@uaslp.mx}
\affiliation{C\'atedras Conacyt, Universidad Aut\'onoma de San Luis Potos\'i, San Luis Potos\'i, 78000 MEXICO}
\affiliation{Coordinaci\'on para la Innovaci\'on y la Aplicaci\'on de la Ciencia y la Tecnolog\'ia, Universidad Aut\'onoma de San Luis Potos\'i,San Luis Potos\'i, 78000 MEXICO}

\date{\today}

\begin{abstract}
We present a quantum model of space charge limited current transport inside trap-free solids with planar geometry in the mean field approximation. We use a simple transformation which allows us to find the exact analytical solution for the steady state current case. We use our approach to find a Mott-Gurney like behavior and the mobility for single charge carriers in the quantum regime in solids.
\end{abstract}


\maketitle

\section{Introduction}
\label{sec:introduction}
The motion of charged particles accelerated across a
gap is of wide interest in fields such as high power
diodes and vacuum microelectronics. In 1910 Child and Langmuir \citep{child,lang} first
studied the space charge limited emission for two infinite parallel plane electrodes
at fixed voltage $V_0$ in vacuum separated by a distance $L$. The one dimensional Child-Langmuir law is given by
\begin{equation}
J=\frac{4\epsilon_0}{9L^2}\sqrt{\frac{2e}{m}}V_0^{3/2},
\label{eqin01}
\end{equation}
where $e$ and $m$ are the charge and mass of the electron, respectively, $\epsilon_0$ is the permittivity of free space.
Since the derivation of this fundamental law many important and useful variations on the classical Child-Langmuir law have been investigated to account for special geometries \citep{lang1,lang2,page}, relativistic electron energies \citep{jory}, non zero initial electron velocities \citep{lang3,jaffe}, quantum mechanical effects \citep{lau,ang,biswas}, nonzero electric field at the cathode surface \citep{barbour}, and slow varying charge density \citep{gg}. \\
The corresponding space charge limited current in a trap-free solid is known as the Mott-Gurney law and is given by \cite{mg}
\begin{equation}
J=\frac{9}{8}\mu\epsilon\frac{V_0^2}{L^3},
\label{eqin01}
\end{equation}
where $\epsilon$ is the permittivity of the solid, and $\mu$ is the electron mobility.\\ With the rapid progress in nanoscale fabrication, nanosize gaps between metallic electrodes have been successfully fabricated. \cite{stein} Thus, there is a renewed interest in the study and application of the Mott-Gurney law in many novel devices, such as graphene oxide sheets, \cite{saiful} light emitting diodes, \cite{torr} organic device, \cite{carbone} and polymer transistor. \cite{chao} In comparison to the recent developments of the Child Langmuir law in the quantum regime \cite{lang4,lang5,chowdhury,sudeep}, the one dimensional Mott-Gurney law has not been studied before in the quantum domain. Therefore, it is of interest to study the space charge limited current in solids placed in a nanogap with low voltage where the electron de Broglie wavelength is comparable to the gap spacing.\\
In this article we examine the quantum extension of the Mott-Gurney Law for solids. We present, for one carrier space charge limited devices, a general transformation to extract the analytical solution of the complete set of equations in the quantum regime. We apply our approach to obtain a Mott-Gurney like behavior for the space charged limited current and the mobility for single charge carriers in the quantum domain in solids.\\ The article is organized as follows. Section II reviews the one dimensional classical model of the Mott-Gurney law. In Section III we set up the main equations of our quantum model for the space charge limited current in solids and we apply a simple transformation to solve them for the steady state current case. In Section IV we show our transformation is a solution to Schr\"odinger's equation for the steady current case. In Section V we perform a simple dimensional analysis to verify the scaling of the current density in the quantum regime. The conclusions are summarized in the last section.
\section{Classical model}
It is very well known that the classical theory of space charge limited current for solids between plane parallel electrodes was first given by Mott and Gurney in 1940. Mott and Gurney first studied the space charge limited current for a trap-free semiconductor sandwiched  between two parallel plane electrodes
at fixed voltage $V_0$ separated by a distance $L$. Assuming that the transverse dimensions are greater than the interelectrode spacing and provided that the thermal carriers are negligible in comparison to the injected ones, a space charge limited regime will be attained and the complete set of equations for this problem are given by
\begin{equation}
J=-\rho v 
\label{eq01a}
\end{equation}
\begin{equation}
\frac{d {\it E}}{dx}=\frac{\rho}{\epsilon} 
\label{eq01b}
\end{equation}
\begin{equation}
{\it E}=-\frac{dV}{dx}
\label{eq01c}
\end{equation}
where $\rho$ is the space charge density, $v$ is the charge drift velocity, ${\it E}$ is the electric field, $\epsilon$ is the electrical permittivity of the material and $V$ is the electric potential. In the case that the velocity increases monotonically with the electric field, i.e. $v=-\mu {\it E}$ where $\mu$ is the mobility, then by using Eq. (\ref{eq01a}) we have
\begin{equation}
J=\rho\mu {\it E}
\label{eq02}
\end{equation}
Equation (\ref{eq02}) shows that the current density is proportional to the charge density and the electric field.
Inserting Eq.(\ref{eq01b}) into Eq. (\ref{eq02}) we have
\begin{equation}
J=\epsilon\mu {\it E}\frac{dE}{dx}.
\label{eq03}
\end{equation}
Integrating Eq. (\ref{eq03}) for the steady state current case and applying the following boundary condition $E(0)=0$ we get
\begin{equation}
E(x)=\sqrt{\frac{2Jx}{\mu\epsilon}}
\label{eq04}
\end{equation}
Using Eq. (\ref{eq01c}) we can integrate again to obtain
\begin{equation}
V(x)=\frac{2}{3}\sqrt{\frac{2J}{\mu\epsilon}}x^{3/2}
\label{eq05}
\end{equation}
Squaring both sides of Eq. (\ref{eq05}) we obtain the Mott-Gurney Law
\begin{equation}
J=\frac{9}{8}\frac{\mu\epsilon}{L^3}V_0^2
\label{eq06}
\end{equation}
\\
\section{Quantum model}
Consider the electrons with energy ${\mathcal E}$ being emitted from the cathode into a solid sandwiched  between two parallel plane electrodes
at fixed voltage $V_0$ separated by a distance $L$. The electric field is applied in the direction perpendicular to the surface of the electrode. Then the current charge density is given in the region $0<x<L$ by
\begin{equation}
J=\frac{e\hbar}{2mi}\left(\psi^{\ast}\frac{d\psi}{dx}-\psi\frac{d\psi^{\ast}}{dx}\right)
\label{eq07}
\end{equation}
where $\psi$ is the electron complex wave function and $J$ is the constant electron emission current density. Using the following transformation (the validity of this transformation is given in section (\ref{appendixA}))
\begin{equation}
\psi=\psi^{\ast}\exp\left[\frac{i}{V_0}\int_0^xEdx\right]
\label{eq08}
\end{equation}
and inserting Eq. (\ref{eq08}) into Eq. (\ref{eq07}) we end up with
\begin{equation}
J=\frac{e\hbar}{2mV_0}\psi^{\ast}\psi E
\label{eq09}
\end{equation}
Comparing equation (\ref{eq09}) with equation (\ref{eq02}) we conclude that the {\it quantum} mobility of the charge carriers is given by
\begin{equation}
\mu_Q=\frac{\hbar}{2mV_0}
\label{eq10}
\end{equation} 
Writing down the applied voltage as $V_0=-(\mu_C-\mu_E)/e$, we can write the quantum mobility in terms of the scattering time of the charge carries, i.e. $\mu_Q=e\tau/m$, where the scattering time is given by
\begin{equation}
\tau=\frac{\hbar}{2(\mu_E-\mu_C)}.
\label{eq11}
\end{equation}
where $\mu_C$ and $\mu_E$ refer to the chemical potential at the collector and injection electrodes, respectively.\\
Using Gauss's law for the electric field
\begin{equation}
\frac{dE}{dx}=\frac{e\psi^{\ast}\psi}{\epsilon}
\label{eq12}
\end{equation}
we can eliminate $|\psi|^2$ from equation (\ref{eq09}) to have
\begin{equation}
J=\mu_Q\epsilon E\frac{dE}{dx}
\label{eq13}
\end{equation}
Integrating Eq. (\ref{eq13}) we have
\begin{equation}
J(x+x_0)=\mu_Q\epsilon \frac{E^2}{2}
\label{eq14}
\end{equation}
where $x_0$ is a constant. Since at $x=0$, $E=J/\mu_Qe|\psi(0)|^2$, we have
\begin{equation}
x_0=\frac{\epsilon J}{2\mu_Qe^2|\psi(0)|^4}
\label{eq15}
\end{equation}
It follows that the electric field intensity is given by
\begin{equation}
E=\sqrt{\frac{2J}{\mu_Q\epsilon}(x+x_0)}
\label{eq16}
\end{equation}
and the potential drop across the solid is
\begin{equation}
V_0=\frac{2}{3}\sqrt{\frac{2J}{\mu_Q\epsilon}}\left[(L+x_0)^{3/2}-x_0^{3/2}\right]
\label{eq17}
\end{equation}
For small applied voltages we have $x_0<<L$, so that the current density given by Eq.(\ref{eq17}) is 
\begin{equation}
J=\frac{9}{8}\mu_Q\epsilon\frac{V_0^2}{L^3}
\label{eq18}
\end{equation}
Equation (\ref{eq18}) is the Mott-Gurney law but with the quantum mobility. Substituting Eq.(\ref{eq10}) into Eq. (\ref{eq18}) we have
\begin{equation}
J=\frac{9}{16}\epsilon\frac{\hbar V_0}{mL^3}
\label{eq19}
\end{equation}
Interestingly, we see from Eq.(\ref{eq19}) that the charge current density follows a linear relationship with the applied voltage in the quantum domain. Our model then shows that at low voltage the conduction is Ohmic. For large voltage, i.e. when space-charges are unimportant, then $x_0>>L$ and we obtain
\begin{equation}
J=\frac{\hbar e}{2mL}|\psi(0)|^2.
\label{eq19a}
\end{equation}
\section{Validity of the transformation}
\label{appendixA}
In the mean field quantum formalism we need to solve the coupled Schr\"odinger-Poisson equations in the Hartree approximation
\begin{equation}
-\frac{\hbar^2}{2m}\frac{d^2\psi}{dx^2}+U(x)\psi={\cal E}\psi 
\label{ap01} 
\end{equation}
\begin{equation}
\frac{d^2V}{dx^2}=\frac{e}{\epsilon}\psi^{\ast}\psi
\label{ap02}
\end{equation}
where ${\cal E}$ is the total energy of the electron and $U$ the potential energy. In this model it is assumed that there is a steady current flowing across the solid with current density given by $J=e\hbar(\psi^{\ast}\psi^{\prime}-\psi\psi^{\ast\prime})/2mi$. Applying the following transformation $\psi=\psi^{\ast}e^{i\phi(x)}$ to equation (\ref{ap01}) we find that the Schr\"odinger equation transforms into
\begin{equation}
{\cal E}\psi^{\ast}=-\frac{\hbar^2}{2m}\left[\psi^{\ast\prime\prime}+2i\phi^{\prime}\psi^{\ast\prime}+i\psi^{\ast}\phi^{\prime\prime}-\psi^{\ast}\phi^{\prime 2}\right]+U\psi^{\ast}.
\label{ap03}
\end{equation}
Applying the transformation to the charge current density we get $J=e\hbar\psi^{\ast}\psi\phi^{\prime}/2m$ and using the continuity equation we get the following identity
\begin{equation}
\frac{dJ}{dx}=-i\psi^{\ast}e^{i\phi}\left[2i\phi^{\prime}\psi^{\ast\prime}+i\psi^{\ast}\phi^{\prime\prime}-\psi^{\ast}\phi^{\prime 2}\right]=0.
\label{ap04}
\end{equation}
Using equation (\ref{ap04}) and the fact that $H\psi^{\ast}={\cal E}\psi^{\ast}$, where $H$ is the Hamiltonian of the system, we see that the transformation $\psi=\psi^{\ast}e^{i\phi(x)}$ is a solution to Schr\"odinger's equation for the steady current case.\\
Since Poisson's equation is proportional to $|\psi|^2$, it is unaffected by our transformation. 

\section{Dimensional Analysis}
To verify the scaling of the current density in the quantum regime we go to the two relevant equations in our derivation which are Gauss's law (Eq. (\ref{eq12}))  together with the current equation (Eq. (\ref{eq07})). Redefining $\tilde\psi=\psi/\sqrt{\epsilon}$ it follows that Gauss's law no longer depends on $\epsilon$ while the current charge density is given by
\begin{equation}
J=\epsilon\frac{e\hbar}{2mi}\left(\tilde\psi^{\ast}\frac{d\tilde\psi}{dx}-\tilde\psi\frac{d\tilde\psi^{\ast}}{dx}\right)
\label{eq20}
\end{equation}
We can now apply the standard dimensional analysis, which is to say that the charge current density can be expressed as
\begin{equation}
J=C_0\epsilon V_0^{\alpha}L^{\delta_1}\hbar^{\delta_2}m^{\delta_3}e^{\delta_4}
\label{eq21}
\end{equation}
where $C_0$ is a dimensionless constant. By expressing all quantities in the M.K.S. units and equating the powers on both sides of Eq.(\ref{eq21}) we obtain the following system of equations
\begin{eqnarray}
\alpha-\delta_4 &=& 1 \\
2\alpha+2\delta_2+\delta_1 &=& 1 \\
\alpha+\delta_2+\delta_3 &=& 1 \\
\delta_4-\delta_2-3\alpha &=& -4  
\label{eq22}
\end{eqnarray} 
The solution to the above system of equations in terms of $\alpha$, the exponent of $V_0$, is given by $\delta_1=2\alpha-5$, $\delta_2=3-2\alpha$, $\delta_3=\alpha-2$ and $\delta_4=\alpha-1$. It follows that the current is given by
\begin{equation}
J=C_0\epsilon V_0^{\alpha}L^{2\alpha-5}\hbar^{3-2\alpha}m^{\alpha-2}e^{\alpha-1}
\label{eq23}
\end{equation}
If $\alpha=1$ in Eq.(\ref{eq23}) we get 
\begin{equation}
J\propto\epsilon \frac{\hbar V_0}{mL^3}
\label{eq24}
\end{equation}
Equation (\ref{eq24}) confirms the power law behavior of the Mott-Gurney law in the quantum domain.
\section{Conclusions}
We have developed a one dimensional quantum model of maximum transmitted current on a trap-free solid sandwiched between two planar electrodes with a fixed applied voltage.
An analytical solution for the space charge limited current in the quantum domain was presented using a simple transformation. It was shown using standard dimensional analysis that the power law obtained is consistent with the solution of our quantum model. 
The method presented can be of help in developing a quantum theory of charge transport in solids for the non trap-free case. 
\section{Acknowledgments}
This work was supported by the program ``C\'atedras CONACYT" and by the project ``Centro Mexicano de Innovaci\'on en Energ\'ia Solar" from Fondo Sectorial CONACYT-Secretar\'ia de Energ\'ia-Sustentabilidad Energ\'etica.\\


\begin{thebibliography}{99}
\bibitem{child} C.D. Child, ``Discharge from hot CaO", Phys. Rev. {\bf 32}, 492-511 (1911) 
\bibitem{lang} I. Langmuir, ``The effect of space charge and residual gases on thermionic currents in high vacuum", Phys. Rev. {\bf 2}, 450-486 (1913)
\bibitem{lang1} I. Langmuir and K.B. Blodgett, ``Currents limited by space charge between coaxial cylinders", Phys. Rev. {\bf 22}, 347-356 (1923)
\bibitem{lang2} I. Langmuir and K.B. Blodgett, ``Currents limited by space charge between concentric spheres", Phys. Rev. {\bf 24}, 49-59 (1924)
\bibitem{page} L. Page and N.I. Adams, Jr., ``Space charge between coaxial cylinders", Phys. Rev. {\bf 68}, 126-129 (1945)
\bibitem{jory} H. R. Jory and A.W. Trivelpiece, ``Exact relativistic solution for the one dimensional diode", J. Appl. Phys. {\bf 40}, 3924-3926 (1969) 
\bibitem{lang3} I. Langmuir, ``The effect of space charge and initial velocities on the potential distribution and thermionic current between parallel plane electrodes", Phys. Rev. {\bf 21}, 419-435 (1923)
\bibitem{jaffe} G. Jaff\'e, ``On the currents carried by electrons of uniform initial velocity", Phys. Rev. {\bf 65}, 91-98 (1944)
\bibitem{lau} Y.Y. Lau, D. Chernin, D.G. Colombant, and P.T. Ho, ``Quantum extension of Child-Langmuir law", Phys. Rev. Lett. {\bf 66}, 1446-1449 (1991)
\bibitem{ang} L.K. Ang, T.J.T. Kwan, and Y.Y. Lau, ``New scaling of Child-Langmuir law in the quantum regime", Phys. Rev. Lett. {\bf 91}, 208303-1-208303-4 (2003)
\bibitem{biswas} Debabrata Biswas and Raghwendra Kumar, ``The Child-Langmuir law in the quantum domain", EPL, {\bf 102}, 58002 (2013)
\bibitem{barbour} J.P. Barbour, W.W. Dolan, J.K. Trolan, E.E. Martin, and W.P. Dyke, ``Space-charge effects in field emission", Phys. Rev. {\bf 92}, 45-51 (1953)
\bibitem{gg} G. Gonz\'alez and F.J. Gonz\'alez Orozco, ``Electron dynamics inside a vacuum tube diode through linear differential equations", J. Plasma Physics {\bf 80}, 247-254 (2014)
\bibitem{mg} N.F. Mott and R. W. Gurney, \textit{Electronic processes in ionic crystals},(Oxford University Press, 1940)
\bibitem{stein} P. Steinmann and J.M.R. Weaver, ``Nanometer-scale gaps between metallic electrodes fabricated using a statistical alignment technique", Appl. Phys. Lett., {\bf 86}, 063104 (2005)
\bibitem{saiful} D. Joung, A. Chunder, L. Zhai and S.I. Khondaker, ``Space charge limited conduction with exponential trap distribution in reduced graphene oxide sheets", App. Phys. Lett., {\bf 97} 093105 (2010)
\bibitem{torr} F. Torricelli, D. Zappa and L. Colalongo, ``Space-charge-limited current in organic light emitting diodes", Appl. Phys. Lett., {\bf 96}, 113304 (2010)
\bibitem{carbone} A. Carbone, C. Pennetta and L. Reggiani, ``Trapping-detrapping fluctuations in organic space-charge layers", Appl. Phys. Lett., {\bf 95} 233303 (2009)
\bibitem{chao} Y.C. Chao, H.F. Menga and S.F. Horng, ``Polymer space-charge-limited transistor", Appl. Phys. Lett. {\bf 88}, 223510 (2006)
\bibitem{lang4} L.K. Ang, W.S. Koh, Y.Y. Lau and T.J.T. Kwan, ``Space-charge-limited flows in the quantum regime", Phys. Plasma {\bf 13}, 056701 (2006)
\bibitem{lang5} W.S. Koh and L.K. Ang, ``Quantum model of space-charge-limited field emission in a nanogap", Nanotechnology, {\bf 19}, 235402 (2008)
\bibitem{chowdhury} Sudeep Bhattacharjee and Tathagata Chowdhury, ``Experimental investigation of transition from Fowler-Nordheim field emission to space-charge-limited flows in a nanogap", Appl. Phys. Lett., {\bf 95}, 061501 (2009)
\bibitem{sudeep} Sudeep Bhattacharjee, Adish Vartak and Victor Mukherjee, ``Experimental study of space-charge-limited flows in a nanogap", Appl. Phys. Lett., {\bf 92}, 191503 (2008) 
\end{thebibliography}
\end{document}